\documentstyle[12pt,epsf]{article}
\begin{document}
\baselineskip=19pt
\begin{titlepage}
\begin{flushright}
  NIIG-DP-01-2\\[-2mm]
  YITP-01-3\\[-2mm]
  hep-ph/0102006
\end{flushright}

\begin{center}
\vspace*{1.5cm}
  
{\large\bf Low-energy constraints from unification of matter
  multiplets}
\vspace{1cm}
  
A.~Kageyama\footnote{E-mail: atsushi@muse.hep.sc.niigata-u.ac.jp},
M.~Tanimoto\footnote{E-mail: tanimoto@muse.hep.sc.niigata-u.ac.jp} and
K.~Yoshioka\footnote{E-mail: yoshioka@yukawa.kyoto-u.ac.jp}
\vspace{5mm}

$^{1,2}${\it Department of Physics, Niigata University,
  Niigata 950-2181, Japan}\\
$^3${\it Yukawa Institute for Theoretical Physics, Kyoto University, 
  Kyoto 606-8502, Japan}
\vspace{1.5cm}
  
\begin{abstract}
  We study the low-energy consequences of the mass and mixing angle
  relations in grand unified theories (GUT), which follow from an
  assumption that some quarks and leptons are placed in the same GUT
  multiplets. This assumption is a simple extension of that for the
  well-known bottom/tau mass ratio, which is one of the most
  successful predictions of grand unification of matter multiplets. We
  show that imposing the GUT relations leads naturally to a limited
  parameter space from the large lepton mixing between the second and
  third generations.
\end{abstract}
\end{center}
\end{titlepage}

\renewcommand{\thefootnote}{\fnsymbol{footnote}}
\setcounter{footnote}{0}

\section{Introduction}

Exploring the origin of the observed fermion masses and mixing angles 
has been one of the most important issues in particle physics. Despite
the fact that there is apparent hierarchical structure among the
masses and mixing angles, no completely definite answer to this
problem has been found. Moreover, the recent neutrino experiments
indicate that neutrinos have non-zero masses and a rather different
mixing pattern than the quark part. Motivated by these facts, many
models have been proposed which try to account for the large lepton
mixing angles in the frameworks beyond the Standard Model.

In these attempts, the desired masses and mixing angles are usually 
fixed by taking parameters in the models as suitable values. In some 
cases, however, dynamical assumptions or symmetry arguments can lead 
relations among the observables. For example, with the $SU(3)$ flavor 
symmetry of the light quarks, the well-known Gell-Mann--Okubo mass 
formula is derived between the octet baryon masses~\cite{GO}. The 
$SU(3)$ symmetry is approximately well valid below the QCD scale and 
the mass formula is in good agreement with the experimental 
values. Another example is the bottom-tau mass equality in $SU(5)$ 
grand unified theories (GUT). The $SU(5)$ gauge symmetry connects the 
bottom quark and tau lepton masses and reproduces the correct
low-energy value, taking into account the renormalization-group (RG)
running effect from the GUT scale down to the weak
scale~\cite{btau,btau2}. In this way, the predictions of such
experimentally well-working relations have convinced us of the
relevance of new symmetries and conjectures with which the relations
hold.

Recently, such a kind of new relation between the quark and lepton 
mixing angles has been proposed in grand unified 
theories~\cite{BKY}. This mixing angle relation is derived from an 
assumption for multiplet structure of quarks and leptons, and
naturally incorporated in unification scenarios. It has been shown
(and as we will see below) that the relation can also be regarded as a
straightforward extension of the bottom-tau mass equality, and it
predicts large lepton mixing between the second and third generations
at the GUT scale.

In this paper, we analyze the low-energy consequences of the above GUT
relations: the bottom-tau mass equality and the mixing angle
relation. We assume the minimal supersymmetric standard model (MSSM)
with right-handed neutrinos below the GUT scale. With the relations as
GUT boundary conditions for the MSSM parameters, we will find new
constraints on the parameter space and the intermediate scale where
the right-handed neutrinos are decoupled.

In the next section, we first review the derivation of the mixing
angle relation proposed in~\cite{BKY} and discuss several its
implications for unification scenarios. We perform the RG analyses in
section 3. It is found that the low-energy value of the bottom-tau
mass ratio becomes more reasonable than the MSSM, while new
constraints generally appear in the case of large Yukawa couplings. In
section 4, we summarize our results and comment on some possible
modifications of the relations.

\section{Mixing angle relations}
\setcounter{equation}{0}

Unification hypotheses for quark and lepton multiplets can lead to
some relations among their Yukawa couplings, that is, masses and
mixing angles. A well-known example of this approach is the down-quark
and charged-lepton mass ratios in GUTs. In this section, we introduce
a relation between the quark and lepton mixing angles and discuss its 
implication in GUT frameworks.

We first consider the mass matrices only for the second and third
generations. For simplicity, we go into the basis in which the
up-quark mass matrix is diagonalized. The quark mass matrices $M_u$
and $M_d$ at the GUT scale are generally written as follows;
\begin{eqnarray}
  M_u &=& 
  \pmatrix{
    m_c & \cr 
    & m_t \cr}, \nonumber\\
  M_d &=& 
  \pmatrix{
    V_{cs} & V_{cb} \cr
    V_{ts} & V_{tb} \cr}
  \pmatrix{
    m_s & \cr
    & m_b \cr}
  \pmatrix{
    V_{\mu 2} & V_{\mu 3} \cr
    V_{\tau 2} & V_{\tau 3} \cr},
  \label{ud}
\end{eqnarray}
where $m_i$ denote the quark mass eigenvalues and $V_{ij}$ the mixing
matrix elements of quarks and leptons. We have assumed that the lepton
mixing is mainly controlled by the charged-lepton mass matrix, that
is, neutrinos have small mixing angles (and probably hierarchical mass
eigenvalues). It is consequently found that the down-quark mass matrix
$M_d$ is diagonalized by the quark and lepton mixing matrices as
above, since at tree level the charged-lepton mass matrix is the
transpose of $M_d$ in GUT frameworks. For the lighter families, some
modification of tree-level mass predictions may be required to be
consistent with the experimental values. However it does not affect
the following discussion because of their tiny Yukawa couplings.

We first impose a natural assumption that the up-quark mass matrix has
a hierarchical form, which is commonly seen in the literature. This
assumption combined with Yukawa unification may also ensure the
hierarchical structure of couplings in the neutrino sector as stated
above. Now we suppose that {\it the right-handed up and down quarks of
  the third generation are contained in a single representation of
  some (GUT) symmetry (and have same couplings)}.
Once this assumption is realized, it is easily found that the mass
matrix element ${M_d}_{23}$ becomes zero because the right-handed up
quarks are rotated such that ${M_u}_{23}$ is set to be zero (i.e.,
$M_u$ is diagonalized). We are thus led to a following equation from
${M_d}_{23}=0$ in (\ref{ud}),
\begin{eqnarray}
  \tan\theta_{\mu\tau} &=& \frac{m_b}{m_s}\, V_{cb},
  \label{relation}
\end{eqnarray}
where $\theta_{\mu\tau}$ is defined by 
$\tan\theta_{\mu\tau}=V_{\mu 3}/V_{\tau 3}$. With this relation at
hand, the lepton 2-3 mixing angle is expressed in terms of the quark
2-3 mixing angle and mass ratio. It is shown in \cite{BKY} that with
the experimental values of the quark masses and mixing angles, the
equation (\ref{relation}) is really consistent with the large lepton
2-3 mixing. One of the purposes of this paper is to see whether the
equation is well valid even at low energies. If one includes the first
generation, the above assumption gives another mixing angle relation
involving the Cabibbo angle, which is also in roughly agreement with
the large lepton mixing~\cite{BKY}. It is interesting that a single
assumption for matter multiplet structure leads to several relations
consistent with experiments. In this paper, however, we focus only
on the heavier families because the inclusion of the first generation
generally involves some ambiguities both experimentally and
theoretically.

It is interesting to notice that the relation (\ref{relation}) can be 
understood as a generalization of the mass relation between the bottom
quark and tau lepton. The bottom-tau mass relation holds when one
requires of the Standard Model that the down quark and charged lepton
of the third generation belong to a single multiplet. Note that this
requirement is very similar to the one which leads to the mixing angle
relation (\ref{relation}). As for the bottom-tau mass equality, the
requirement results in the $SU(5)$ GUT symmetry beyond the Standard
Model. The fact that the mass relation is experimentally
well-supported at low energies is one of the great motivations to
investigate $SU(5)$ models. In the present case, the assumption that
the right-handed up and down quarks come from a single multiplet
clearly requires some symmetries beyond the $SU(5)$, such as the
left-right symmetry $SU(2)_R$,\footnote{The left-handed up and down
  quarks already belong to the same multiplets of the Standard Model
  gauge group $SU(2)_L$.}
$SO(10)$, and $E_6$ gauge symmetries. As mentioned above, the mixing
angle relation is consistent with the experimental results at the GUT
scale. This fact may also be one of the convincing proofs of the
existence of larger symmetries at high energies.

Before proceeding to low-energy RG analyses, we here discuss some
implications of the assumption in GUT multiplet structures. In the
following, we use the $SU(5)$ GUT language, for simplicity. In $SU(5)$ 
GUTs, all quarks and leptons are assigned into $10$ and $5^*$ 
representations of $SU(5)$. The left-handed quarks belong to 
$10$-dimensional representation fields and the left-handed lepton are 
in $5^*$ fields. As a result, the quark mixing angles are determined
only by the structure of $10$-dimensional fields, and on the other
hand, the lepton mixing only by that of $5^*$-representation
fields. One of the simple ways to achieve the small quark mixing is
that hierarchical suppression factors are attached to $10$-dimensional
fields. This can be realized in various dynamical ways~\cite{10dim},
for example, with a charge assignment of $U(1)$ flavor symmetries. One
of the assumptions we adopt in the above, that is, a hierarchy in the
up-quark masses, is thus established. Interestingly, that also explains
the larger hierarchy among the observed up-quark masses than those of
the down quarks and the charged leptons. On the other hand, it has
been experimentally confirmed that there exists large generation
mixing in the lepton side~\cite{superK}. An interesting idea to
explain this observation within GUT frameworks is to mix the
$5^*$-representation fields. That is, by rotating $5^*$'s, one can
obtain large lepton mixing while the quark mixing angles remain
small. It has been shown that this mechanism is indeed incorporated
naturally in the GUT frameworks based on $SU(5)$~\cite{SY},
$SO(10)$~\cite{NY}, and $E_6$~\cite{BK} gauge groups.

To have proper values of the mixing angles, we need an implicit
requirement that ${M_d}_{32}$ is of the same order of
${M_d}_{33}$. This is just a condition which associates the quark
mixing angle with the lepton one. However it should be noted that the
equation (\ref{relation}) itself holds without any such additional
assumptions, besides that for matter multiplet structure. Since we
suppose a hierarchical form of $10$-dimensional fields structure, one
might wonder that the $5^*$ fields also have a hierarchy and the large
mixing angle condition, ${M_d}_{32}\sim {M_d}_{33}$, may not be
realized in GUT frameworks. The $5^*$-rotating mechanism now gives a
simple solution to this problem. That is, one can satisfy the
condition by suitably twisting the second and/or third generation
$5^*$ fields, preserving the small quark mixing angles. In the
$SO(10)$ model~\cite{NY}, the original second generation $5^*$ is
replaced by another $5^*$ in the extra matter fields so that the large
mixing condition is fulfilled. Moreover, no particular assumption is
imposed on the third-generation fields and thereby they have the
third-generation $10_3$ and $5^*_3$ coming from a single field $16_3$
in $SO(10)$. As a result, the mixing angle relation (\ref{relation})
certainly follows. For another example, in the $E_6$ model~\cite{BK}
which utilizes a similar $5^*$-rotating mechanism, the situation is a
bit different. In a fundamental representation $27$ of $E_6$, the
$5^*$ representation of $SU(5)$ appears twice and hence one naturally
has a possibility to choose the low-energy $5^*$ candidates without
any extra matter fields. This is an interesting feature of $E_6$ GUT
models. In particular, in Ref.~\cite{BK}, the third generation $5^*_3$
is exchanged with the inherent $5^{*'}_3$ in the same $27$-dimensional
field. Since $10_3$ and $5^{*'}_3$ have the same couplings connected
by the $E_6$ gauge symmetry, the relation (\ref{relation}) also holds
in this case.

After all, we find that the mixing angle relation is one of the
significant predictions of GUT models with $5^*$-rotations, which
ensure the lepton large mixing. But we would like to emphasize that
the relation generally holds under only one assumption for the matter
multiplet structure and is independent of any details of
models. Moreover it has much more generality and may be valid without
$5^*$-rotations. For example, the relation also follows in an $SO(10)$
GUT model~\cite{ABB} and with a special ansatz of mass
texture~\cite{NNI}. To check the validity of the relation at the
electroweak scale will be an important probe for internal family
structure at high energies.

\section{RG evolution of the GUT relations}
\setcounter{equation}{0}

We have shown in the previous section that the simple hypotheses for
matter multiplet structure, which imply the beyond the Standard Model,
lead to two GUT relations: the bottom-tau mass equality and the mixing
angle relation (\ref{relation}). In this section, we study the
low-energy consequences of the relations by assuming that the
low-energy effective theory below the GUT scale is the MSSM with
right-handed neutrinos. We are not concerned with particular
mechanisms in GUT models which actually induce the relations. Instead,
the relations are treated as the boundary conditions for Yukawa
couplings of the MSSM\@. According to the results in the previous
section, an appropriate choice for the boundary values of the second
and third generation Yukawa couplings is
\begin{eqnarray}
  Y_u|_{\rm GUT} \;=\; c_u\pmatrix{
    & f \cr
    & 1 \cr},\quad 
  Y_d|_{\rm GUT} \;=\; c_d\pmatrix{
      & f \cr
    h & 1 \cr},\quad 
  Y_e|_{\rm GUT} \;=\; c_d\pmatrix{
      & h \cr
      & 1 \cr},
  \label{bc}
\end{eqnarray}
where $f$ and $h$ are the small ($O(10^{-(1-2)})$) and $O(1)$
constants, respectively. We have explicitly denoted only the relevant
matrix elements which are responsible for the mass and mixing angle
relations. The other elements are negligibly small in each matrix and
do not affect the RG analyses in what follows. There are two essential
points in the above form of boundary conditions. The one is that
the second column of $Y_d$ is the same as the second row of $Y_e$. The
other is the second rows of $Y_u$ and $Y_d$ are proportional to each
other. The former condition gives the bottom-tau mass equality, and
the latter one, together with the former, produces the mixing angle
relation. In fact, this is a weaker assumption than that we have
argued in section 2 in a sense that the proportionality constants
$c_u$ and $c_d$ generally take different values. However it is
certainly a sufficient condition to obtain the mixing angle
relation. The constants $c_u$ and $c_d$ are model-dependent and fix
the ratio of two vacuum expectation values of the Higgs doublets in
the MSSM\@. For example, in the $E_6$ model with $5^*$ rotations
discussed before, a hierarchy between the two constants, $c_u\gg c_d$,
is preferred and actually realized assuming relevant Higgs couplings
at the GUT scale.

To estimate the evolution of the relations down to low energies, it is
useful to define the following two quantities;
\begin{eqnarray}
  R &\equiv& \frac{m_b}{m_\tau} \;=\; \left(
    \frac{{Y_d}_{32}^2+{Y_d}_{33}^2}{{Y_e}_{23}^2+{Y_e}_{33}^2}
  \right)^{1/2}, \\
  X &\equiv& \frac{V_{cb}}{\tan\theta_{\mu\tau}} \frac{m_b}{m_s} 
  \nonumber \\
  &=& \frac{{Y_d}_{32}^2+{Y_d}_{33}^2}{{Y_d}_{22}{Y_d}_{33}
    -{Y_d}_{32}{Y_d}_{23}}\, \frac{{Y_e}_{33}}{{Y_e}_{23}} 
  \left(\frac{{Y_d}_{22}{Y_d}_{32}+{Y_d}_{23}{Y_d}_{33}}
  {{Y_d}_{32}^2+{Y_d}_{33}^2}-\frac{{Y_u}_{23}}{{Y_u}_{33}} \right).
\end{eqnarray}
The subscripts 2, 3 are the generation indices. At the GUT scale, the
boundary conditions are $R|_{\rm GUT}=X|_{\rm GUT}=1$ due to the GUT
relations we adopt. It is interesting to note that the boundary value
of $X$ is independent of that of the small Yukawa couplings such as
${Y_d}_{22}|_{\rm GUT}$, though the expression of $X$ partly contains
these Yukawa couplings. The relevant couplings for $X|_{\rm GUT}$ are
only those we have described in the matrices (\ref{bc}). With this
freedom at hand, we can tune the mass eigenvalues of the second
generation, for example, by use of the Georgi-Jarlskog
factor~\cite{GJ} coming from non-minimal representations for the Higgs
sector, and supersymmetric loop corrections~\cite{LMP}. Such a detail
of small Yukawa couplings, however, has nothing to do with the RG
analyses and we will safely neglect their effects.

The 1-loop RG equations for $R$ and $X$ read
\begin{eqnarray}
  \frac{dR}{dt} &=& \frac{R}{16\pi^2} \Biggl[{Y_u}_{33}^2
    -\biggl(\frac{{Y_e}_{33}^2}{{Y_e}_{23}^2+{Y_e}_{33}^2}\biggr)
    {Y_\nu}_{33}^2 +3({Y_d}_{32}^2+{Y_d}_{33}^2
      -{Y_e}_{23}^2-{Y_e}_{33}^2) \Biggr. \nonumber \\[2mm]
    &&\Biggl.\qquad -\frac{16}{3}g_3^2+\frac{4}{3}g_1^2 
    \Biggr], 
  \label{R} \\
  \frac{dX}{dt} &=& \frac{X}{16\pi^2} \left[
    2({Y_d}_{32}^2+{Y_d}_{33}^2)+{Y_\nu}_{33}^2 \right],
  \label{X}
\end{eqnarray}
where $t=\ln\mu$ is the renormalization scale, $Y_\nu$ is the neutrino
Yukawa coupling, and $g_{1,3}$ are the gauge coupling constants of the
$U(1)$ hypercharge and $SU(3)$ color gauge groups, respectively. We
have assumed that the neutrino Yukawa matrix has a hierarchical form
similar to the up-quark matrix, at least between the second and third
generations. This assumption is usually expected in GUT frameworks.

We first discuss bottom-tau unification, $R|_{\rm GUT}=1$. It is known
that in the MSSM with right-handed neutrinos, bottom-tau unification
casts severe constraints on the parameter space~\cite{btauMr}. In
particular, there are two parameters which are restricted in the
presence of the tau-neutrino Yukawa coupling. One is $\tan\beta$
which is defined by the ratio of vacuum expectation values of the
up- and down-type Higgs doublets. The other is $M_R$ where the
(third-generation) right-handed neutrinos are decoupled. The essential
point they argue in \cite{btauMr} is that in the RG equation for the
bottom-tau ratio, the effect of tau-neutrino Yukawa coupling cancels
that of the top-quark one. The latter plays a significant role to
produce the proper low-energy value of the bottom-tau ratio in the
case with no neutrino Yukawa couplings. As a consequence, a small
value of $\tan\beta$ (i.e., small bottom and tau Yukawa couplings) is
disfavored, and in addition, a lower scale of $M_R$ is excluded
because the contribution of tau neutrino becomes larger. Several
models have been discussed to ameliorate this problem with extra
matter and gauge dynamics~\cite{btau-sol}.

\begin{figure}[htbp]
\begin{center}
  \leavevmode
  \epsfxsize=10cm \ \epsfbox{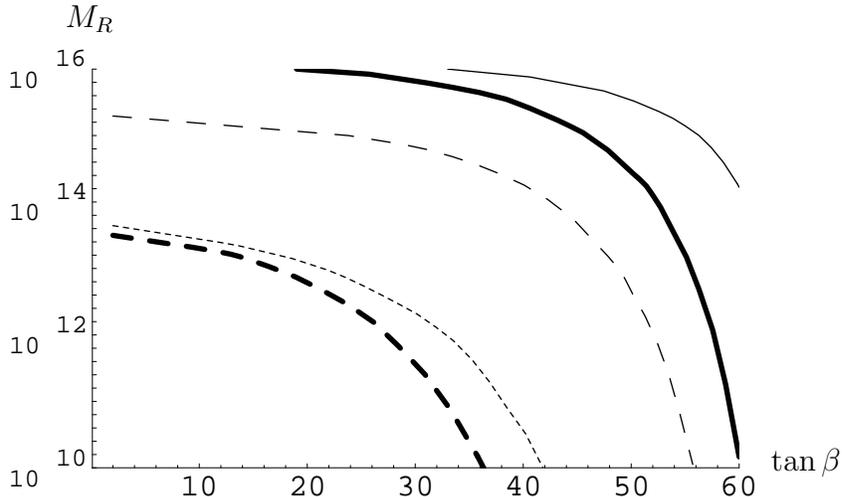}
  \put(5,11){$\tan\beta$}
  \put(-262,179){$M_R$}
  \caption{An illustration of the lower bounds on $M_R$ and
    $\tan\beta$ from bottom-tau unification, with our boundary
    condition (thick) and the MSSM ones (thin) for
    $\alpha_3(M_Z)\equiv g_3^2(M_Z)/4\pi=0.115$ (dotted), 0.12
    (dashed), and 0.125 (solid). The thick dotted curve is on the
    outside of this region. In all cases, ${Y_u}_{33}|_{\rm GUT}=2.5$
    and an upper bound of the bottom-quark mass $m_b^{pole}=5.2$ GeV
    are used.}
  \label{fig:btau}
\end{center}
\end{figure}

In the present situation, the matter contents and couplings are just
the same as that in Ref.~\cite{btauMr}, but the important difference
is in the boundary condition of the Yukawa couplings. We do impose
bottom-tau unification at the GUT scale but some additional $O(1)$ 
couplings are included into the analysis, that is relevant for the
lepton large mixing. It is seen from (\ref{R}) that with our boundary
condition, the cancellation between the top and tau-neutrino Yukawa
couplings is rather reduced even close to the GUT scale, for
${Y_e}_{33}^2/({Y_e}_{23}^2+{Y_e}_{33}^2) \simeq 1/2$. In
Fig.~\ref{fig:btau}, we illustrate typical lower bounds on $\tan\beta$
and $M_R$ from bottom-tau mass unification. In this figure, we take
the top Yukawa coupling and an upper bound for the physical pole mass
of the bottom quark as ${Y_u}_{33}|_{\rm GUT}=2.5$ and
$m_b^{pole}=5.2$ GeV, which may be conservative values in estimating
allowed parameter space. The dashed lines correspond to the case of
the usual MSSM boundary condition (i.e., that of Ref.~\cite{btauMr})
and the solid ones for the present case. For a smaller value of
${Y_u}_{33}$, the low-energy prediction for $R$ is similar to that in
the usual MSSM case. This is because the net effect of $Y_d$ and $Y_e$
in the RG evolution is same in both cases, given the low-energy values
of fermion masses. We find from the figure that the constraint from
bottom-tau unification is weakened and in particular, for small gauge
coupling constants, physically meaningful bounds on $M_R$ disappear
even in case of small $\tan\beta$. One of the interesting influences
of this result is on the lepton flavor violation phenomenon. It is
known that the lepton flavor violation processes induced by the
right-handed neutrinos are largely enhanced with $\tan\beta$. That
combined with the Yukawa matrix form like eq.~(\ref{bc}) is shown to
already exclude a large value of $\tan\beta$~\cite{LFV}. However we
find in the above analysis that the small $\tan\beta$ region is still
available even with bottom-tau unification, unlike the usual MSSM
case. It is interesting that the boundary condition, which leads to
the GUT relations for large lepton mixing, gives at the same time a
simple solution to the bottom-tau unification problem.
 
\begin{figure}[htbp]
\begin{center}
  \leavevmode
  \epsfxsize=10cm \ \epsfbox{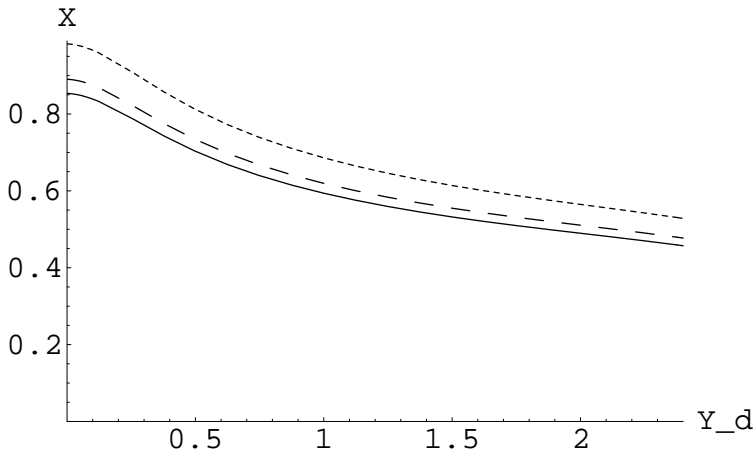}
  \caption{Typical values of the mixing angle ratio $X$ at the weak
    scale. The horizontal axes denotes the value of down-quark Yukawa
    couplings at the GUT scale: 
    $Y_d\equiv {Y_d}_{32}|_{\rm GUT}={Y_d}_{33}|_{\rm GUT}$. The
    solid, dashed, and dotted curves correspond to $M_R=10^{10}$,
    $10^{13}$, and $10^{16}$ GeV, respectively.}
  \label{fig:X}
\end{center}
\end{figure}

\begin{figure}[htbp]
\begin{center}
  \leavevmode
  \epsfxsize=10cm \ \epsfbox{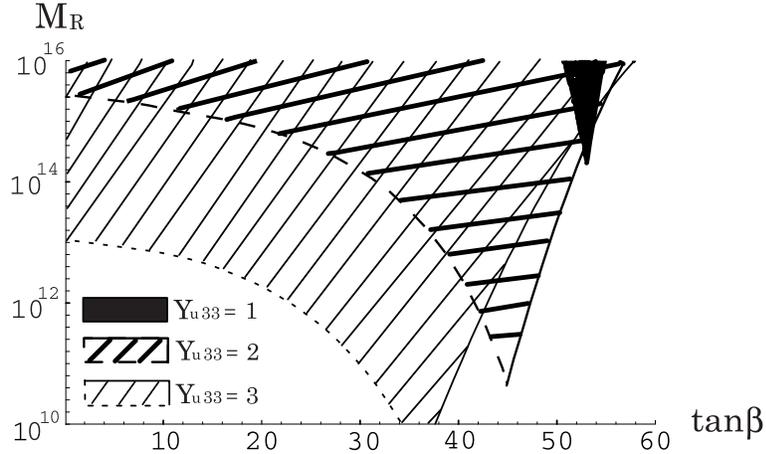}
  \caption{The parameter regions for $(\tan\beta, M_R)$ experimentally
    allowed with the constraints from two GUT relation: $R=X=1$. Three
    different regions are shown for the values of ${Y_u}_{33}=1$, 2,
    and 3. The averaged value of $\alpha_3(M_Z)=0.118$ is used.}
  \label{fig:RX}
\end{center}
\end{figure}

\begin{figure}[htbp]
\begin{center}
  \leavevmode
  \epsfxsize=10cm \ \epsfbox{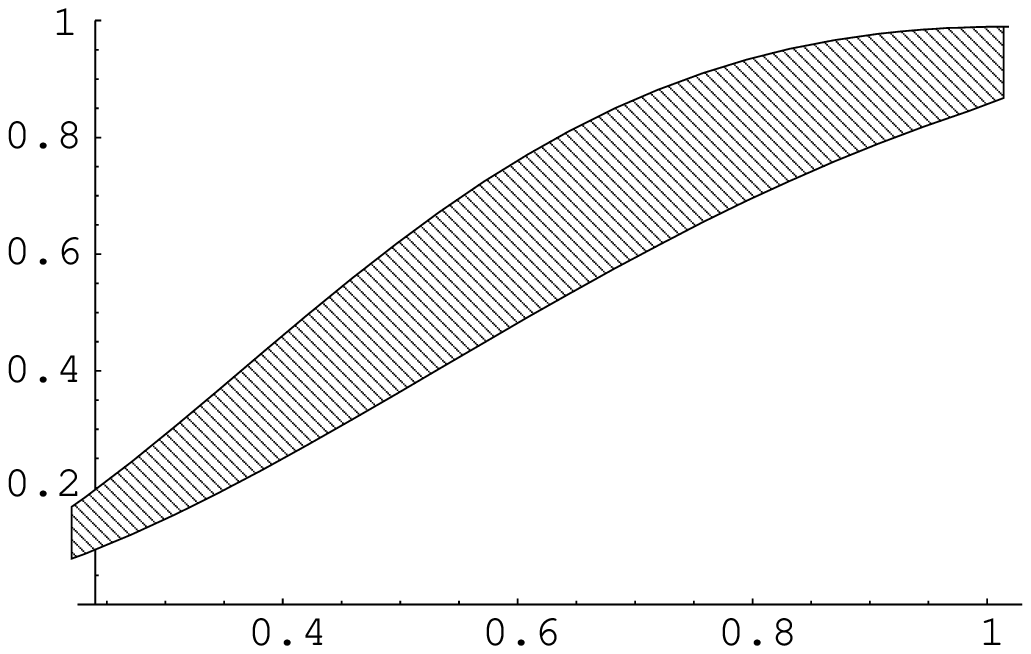}
  \put(6,10){$X$}
  \put(-275,190){$\sin^2 2\theta_{\mu\tau}$}
  \caption{The dependence of the lepton mixing angle 
  $\sin^2 2\theta_{\mu\tau}$ on $X$ with the experimental values of
  the quark masses and mixing angle as an input. $X=1$ corresponds to
  the exact GUT relation (or the case that $X$ is RG-invariant). The
  weak-scale value $m_s=100$ MeV is used, but a smaller value of $m_s$
  tends to give a severer bound on $\sin^2 2\theta_{\mu\tau}$.}
  \label{fig:sin}
\end{center}
\end{figure}

Another GUT relation, the mixing angle relation (\ref{relation}), has
rather different consequences for the low-energy physics. The RG
evolution of $X$ is governed only by the down-quark and neutrino
Yukawa couplings at one-loop level. The low-energy value of $X$ hence
depends on the two parameters, $M_R$ and $\tan\beta$. The intermediate
scale $M_R$ determines the size of the contribution of neutrino Yukawa
couplings, and on the other hand, $\tan\beta$ corresponds to the
down-quark Yukawa contribution, given a fixed value of the top quark
mass. From the RG equation (\ref{X}), we find that if the right-handed
tau neutrino mass $M_R$ is close to the GUT scale and $\tan\beta$ is
small, the mixing angle relation, i.e., $X$, is effectively
RG-invariant. The dependence of the low-energy $X$ on these two
parameters is shown in Fig.~\ref{fig:X}.
We see from this figure that the low-energy value of $X$ has a
relatively mild $M_R$-dependence but is a monotonically decreasing
function of the down-quark Yukawa couplings. Since the gauge coupling
contributions to the RG evolution is very small even at higher-loop
level, the maximal value of $X$ turns out to be 1. As we discussed in
section 2, at the GUT scale, the mixing angle relation is in good
agreement with the experimentally observed large lepton mixing. A
deviation from the exact relation $X=1$, therefore, could exclude a
part of the parameter space in the model, in particular, the large
$\tan\beta$ (i.e., large $Y_d$) region. We perform a numerical
analysis with the 2-loop RG equations and show in Fig.~\ref{fig:RX}
the parameter region allowed by the constraints from two GUT
relations.
We have taken $X(M_Z)>0.8$ as an experimental lower bound of the
low-energy value. This is roughly translated into a lower bound of the
lepton mixing angle, $\sin^2 2\theta_{\mu\tau}>0.9$ (see
Fig.~\ref{fig:sin}), which is indicated by the Superkamiokande
experiment~\cite{superK}. In the whole parameter space, the small
$\tan\beta$ region is excluded from the first relation, the bottom-tau
mass equality. For a smaller value of $\tan\beta$, the bottom quark
becomes heavier beyond the experimental limit (or the tau lepton is
too light). On the other hand, the large $\tan\beta$ region is
eliminated by the second relation, the mixing angle relation, which is
one of the characteristics of the present models. A large value of the
bottom Yukawa coupling reduces the lepton mixing angle during the RG
evolution down to low energies. In this way, the two GUT relations
play complemental roles in obtaining the limits for the model
parameters. In the viewpoint of the top Yukawa coupling, the
complementarity is more clearly seen. That is, with a smaller
${Y_u}_{33}$, the constraint from bottom-tau unification becomes
severer, but that from the mixing angle relation is less important in
the case of top-neutrino Yukawa unification.

We find that only the intermediate $\tan\beta$ region is typically
left in this analysis. There has been a discussion in the MSSM that
this region of $\tan\beta$ is disfavored by the observed value of the
top quark mass, under the bottom-tau unification
assumption~\cite{btau2}. Their arguments, however, depend on a 
particular choice of parameters such as $\alpha_3$, $m_t$, and the
boundary conditions for Yukawa and supersymmetry breaking terms. In
the present models with the boundary conditions (\ref{bc}), different
results should be expected. We leave these analyses, including
possible corrections to the relations, to future investigations.

\section{Summary and discussion}
\setcounter{equation}{0}

In summary, we have studied the low-energy consequences and validity
of the GUT-scale relations among the masses and mixing angles. These
relations follow from the simple hypotheses for multiplet structure of
quarks and leptons. The first relation is the celebrated bottom-tau
mass equality. It is derived with an assumption that the
third-generation down quark and charged lepton belong to a single
multiplet of some symmetry beyond the Standard Model. This requirement
results in the $SU(5)$ grand unification. The second relation we have
discussed is a straightforward extension of the first one. The
assumption in this case is that the right-handed up and down quarks in
the third generation come from a single multiplet. We thus have the
mixing angle relation, which connects the lepton 2-3 mixing angle with
the quark mixing angle and mass ratio. This assumption clearly
suggests the existence of some dynamics beyond the $SU(5)$ gauge
symmetry, for example, higher unification of $SO(10)$ or $E_6$. The
fact that the mixing angle relation is indeed experimentally
well-supported may convince us of such new physics at high
energies. We have also discussed that to rotate $5^*$ representation
fields is one of the simplest ways to achieve the large lepton 2-3
mixing while the quark mixing angle remains small.

To see the low-energy consequences of the GUT-scale relations, we have
performed the renormalization-group analyses in the MSSM with
right-handed neutrinos. We have adopted a particular choice of the
boundary conditions for Yukawa couplings suggested by the
relations. We first shown that the low-energy prediction for the
bottom-tau ratio can be fitted to the experimental value better than
the usual MSSM case. That gives a simple resolution to the unwilling
situation argued in Refs.~\cite{btauMr}. The analyses are deeply
concerned with two model-parameters: the intermediate scale $M_R$ and
$\tan\beta$. From the experimental bounds on the GUT-scale relations,
we have plotted the allowed range for these two parameters. There each
GUT relation eliminates each side (small or large value) of
$\tan\beta$, and the two GUT relations thus play complementary
roles. As a result, only the intermediate range of $\tan\beta$ is
still left available.

There may be some sources of deviations from the exact GUT-scale
relations. First, one could have non-negligible lepton mixing from the
neutrino side. Even if the up-neutrino Yukawa unification is assumed,
a large mixing angle can be predicted depending on the forms of
right-handed neutrino Majorana mass terms (for 
example, \cite{majorana}). That, however, generally requires a
hierarchy in the mass matrices and hence some tuning of
parameters. The threshold corrections at the GUT and supersymmetry
breaking scales also give possible modifications of the
relations. They could become considerably large, in particular, for
the bottom quark mass via the 1-loop diagrams involving supersymmetric
particles, gluino and chargino~\cite{threshold}. However these
corrections have large ambiguities depending on the field contents and
symmetries in the models and are less under our control. Once
supersymmetry is discovered in future experiments, we can definitely
evaluate the corrections for the first time. Note that these
complicated and potentially large supersymmetry contributions do not
modify the mixing angle relation. Deviations from the exactness
$X=1$ is predominated only by the RG evolution discussed in this
paper. Thus the mixing angle relation has an advantage for finding a
clue to high-energy physics in that it is not spoiled by any threshold
effects at intermediate scales. Combined with these factors, more
precise measurements of the quark and lepton mixing angles would probe
higher-scale matter multiplet structure via the GUT-scale relations as
its remnants.

\vspace*{8mm}
\subsection*{Acknowledgments}

We would like to thank to N.~Okamura for useful discussion and
comments. We also thank the organizers and participants of Summer
Institute 2000 held at Yamanashi, Japan, where this work was inspired
by stimulating seminars and discussions.


\end{document}